\documentclass[sigconf]{acmart}
\acmConference[EASE 2026]{The 30th International Conference on Evaluation and Assessment in Software Engineering}{10–13 June, 2026}{Glasgow, Scotland, United Kingdom}

\usepackage{pifont} 
\usepackage{enumitem}

\usepackage{subfigure}
\usepackage{graphicx}
\usepackage{multicol}
\usepackage{multirow}
\usepackage{xcolor,colortbl}
\usepackage[most]{tcolorbox}
\usepackage{textcomp}%
\usepackage{manyfoot}%
\usepackage{threeparttable}
\usepackage{array}
\usepackage{amsmath}
\usepackage{tikz}
\usetikzlibrary{positioning, shapes.geometric, arrows.meta}

\definecolor{mycolor_box}{HTML}{F5FFFA} 
\definecolor{mycolor_title}{HTML}{FFEBCD} 

\AtBeginDocument{%
  }

\setcopyright{acmlicensed}
\copyrightyear{2025}
\acmYear{2025}
\acmDOI{XXXXXXX.XXXXXXX}

\acmISBN{978-1-4503-XXXX-X/18/06}

\copyrightyear{2025}
\acmYear{2025}
\setcopyright{rightsretained}
\acmPrice{}
\acmDOI{10.1145/3756681.3756985}
\acmISBN{979-8-4007-1385-9/25/06}
\begin{document}



\title{\textit{MoEKD}: Mixture-of-Experts Knowledge Distillation for Robust and High-Performing Compressed Code Models}

\author{Md. Abdul Awal}
\affiliation{
  \institution{University of Saskatchewan, Canada}
  \city{}
  \country{}
}
\email{abdul.awal@usask.ca}

\author{Mrigank Rochan}
\affiliation{
  \institution{University of Saskatchewan, Canada}
  \city{}
  \country{}
}
\email{mrochan@cs.usask.ca}

\author{Chanchal K. Roy}
\affiliation{
  \institution{University of Saskatchewan, Canada}
  \city{}
  \country{}
}
\email{chanchal.roy@usask.ca}

\begin{abstract}
Large language models for code have achieved strong performance across diverse software analytics tasks, yet their real-world adoption remains limited by high computational demands, slow inference speeds, significant energy consumption, and environmental impact. Knowledge distillation (KD) offers a practical solution by transferring knowledge from a large model to a smaller and more efficient model. Despite its effectiveness, recent studies show that models distilled from a single source often exhibit degraded adversarial robustness, even when robustness-aware distillation techniques are employed. These observations suggest a fundamental limitation of single-source distillation in simultaneously transferring high-quality and robust knowledge. To overcome this limitation, we propose Mixture of Experts Knowledge Distillation (\textit{MoEKD}), a KD framework that leverages a Mixture of Experts (MoE) architecture to enable more effective and robust knowledge transfer from multiple specialized experts into a compact model. \textit{MoEKD} decomposes the distillation process into expert and router training, aggregation of expert knowledge through a learned routing mechanism, and distillation from the aggregated knowledge. We evaluate \textit{MoEKD} on the vulnerability detection task using CodeBERT and GraphCodeBERT models. Experimental results show that \textit{MoEKD} not only improves adversarial robustness by up to \textbf{35.8\%}, but also enhances predictive performance by up to \textbf{13\%}, compared to state-of-the-art KD baselines, including \textit{Compressor} and \textit{AVATAR}. Furthermore, an ablation study demonstrates that aggregating expert knowledge enables ultra-compact models to maintain competitive performance even when their size is reduced by approximately half. Overall, these results highlight the effectiveness of multi-expert knowledge aggregation in addressing key limitations of existing single-source KD approaches.

\end{abstract}



\maketitle


\section{Introduction}
\label{Intro}
Large language models for code have attracted significant attention in the software engineering community for their ability to support a wide range of tasks, including clone detection \cite{arshad2022codebert}, code summarization \cite{ahmed2024automatic}, vulnerability detection \cite{ding2024vulnerability}, and code search \cite{chen2024code}. Despite these advances, deploying these models on consumer-grade devices and in latency-sensitive applications remains challenging because of high computational costs, slow inference speeds, increased energy consumption, and a considerable carbon footprint \cite{schwartz2020green}. To address these challenges, researchers have widely investigated knowledge distillation (KD) as a model compression technique that transfers knowledge from a large (teacher) model to a smaller (student) model, thereby reducing model size, increasing inference speed, and mitigating environmental impact, while largely preserving task performance \cite{shi2024efficient, shi2024greening, liu2026pioneer, panichella2025metamorphic, chen2025smaller}. However, Panichella et al. \cite{panichella2025metamorphic} empirically demonstrated that knowledge-distilled student models often exhibit reduced robustness and introduced \textit{MORPH}, a KD-based approach that explicitly incorporates robustness into the distillation process. Subsequent work by Awal et al. \cite{awal2025metamorphic} showed that, despite these improvements, student models trained with MORPH remain less robust than their teacher counterparts under adversarial attacks. These findings highlight a fundamental limitation of existing KD approaches that rely on a single teacher model to transfer knowledge to compact student models, a limitation that, to the best of our knowledge, has not been previously examined.

\begin{figure}[htbp]
\centering
\includegraphics[width=6.5cm, height = 7cm]{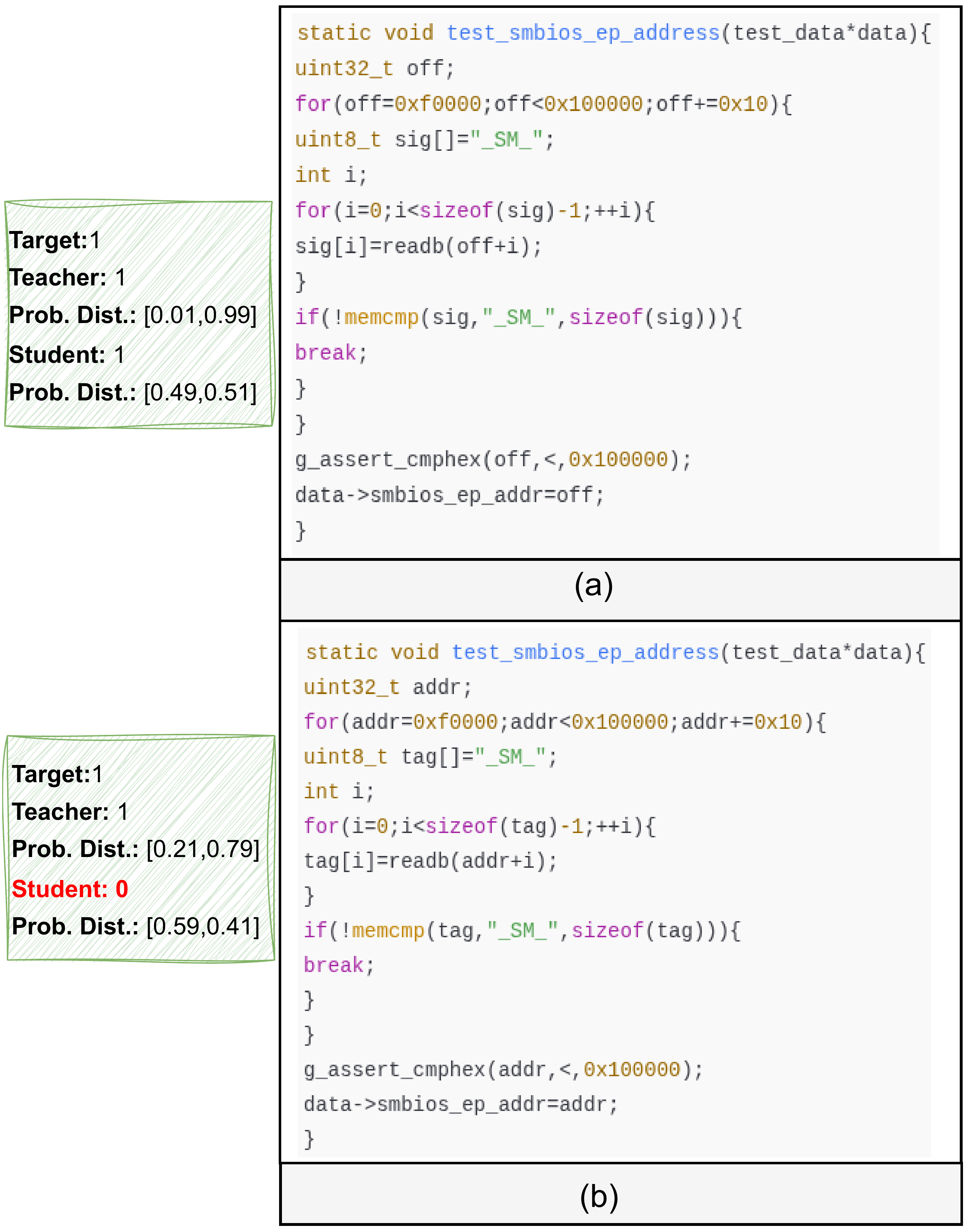}
\caption{A motivating example illustrating limited knowledge transfer in single-teacher distillation under minor adversarial perturbations.} 
\label{MotivationMoEKD}
\end{figure}

To better understand this limitation, it is important to examine knowledge transfer at the level of predictive confidence rather than focusing solely on the final prediction. As illustrated in Figure~\ref{MotivationMoEKD}, a teacher model and its distilled student assign the same vulnerability label to a code fragment under standard conditions; however, the teacher exhibits substantially higher confidence (\textbf{0.99 vs. 0.51}), reflecting a more stable decision boundary. Under a minor adversarial perturbation (e.g., renaming identifiers: \textit{off} $\rightarrow$ \textit{addr} and
\textit{sig} $\rightarrow$ \textit{tag}), the teacher’s confidence decreases but remains above the decision threshold, preserving the correct prediction, whereas the student’s confidence drops further, leading to misclassification. This example suggests that knowledge transferred from a single teacher may be limited in scope and insufficient to adequately represent diverse or challenging decision boundaries, especially those sensitive to adversarial perturbations, underscoring the need for richer, more informative supervisory knowledge during distillation.

Motivated by this observation, we posit that more effective knowledge distillation can be achieved by aggregating complementary knowledge from multiple specialized teachers rather than relying on a single source. This perspective is analogous to higher education, where students develop a comprehensive understanding of an entire curriculum by learning from multiple instructors, each specializing in a particular subject area, rather than from a single generalist. Such specialization enables each teacher to provide deeper, more focused knowledge within their domain, allowing students to develop a more comprehensive and robust understanding. Extending this intuition to the knowledge distillation context, supervision from multiple specialized expert models can provide richer and more diverse knowledge than supervision from a single teacher. By aggregating knowledge from these complementary experts, the student model is exposed to a broader spectrum of task-relevant information that may not be fully captured by any single teacher. A \textbf{Mixture of Experts} (MoE) \cite{jacobs1991adaptive,yuksel2012twenty} framework provides a principled abstraction for this setting by enabling the systematic extraction and integration of expert knowledge during distillation. Building on this rationale, we propose \textbf{Mixture of Experts Knowledge Distillation} (\textit{MoEKD}), a distillation framework that leverages a Mixture of Experts architecture to more effectively transfer knowledge from multiple specialized teachers to a compact student model.

At a high level, \textit{MoEKD} decomposes the knowledge distillation process into three key components: \textit{training phase}, \textit{expert aggregation and logits fusion}, and \textit{knowledge distillation}. In the training phase, multiple expert models are trained independently to specialize in distinct regions of the input space, and a router model is trained to select the most relevant experts for each input. During expert aggregation and logits fusion, a routing mechanism selectively combines the outputs of the top-\textit{k} relevant experts to generate unified supervisory knowledge. This aggregated knowledge is subsequently transferred to a compact student model through a dedicated distillation process.

We empirically evaluate the effectiveness of \textit{MoEKD} in enhancing both predictive performance and adversarial robustness of compressed student models, using two popular language models for code, CodeBERT \cite{feng2020codebert} and GraphCodeBERT \cite{guo2020graphcodebert}, for the vulnerability detection task. Adversarial robustness is measured using three widely studied identifier renaming attacks: ALERT \cite{yang2022natural}, Metropolis Hastings Modifier (MHM) \cite{zhang2020generating}, and WIR-Random \cite{zeng2022extensive}. The experimental results indicate that \textit{MoEKD} increases the predictive performance of student models by up to \textbf{13.0\%} compared to state-of-the-art (SOTA) baselines, including \textit{Compressor} \cite{shi2022compressing} and \textit{AVATAR} \cite{shi2024greening}. Regarding adversarial robustness, \textit{MoEKD} achieves improvements of up to \textbf{35.8\%} over these baselines. Furthermore, an ablation study demonstrates that the high-quality knowledge aggregated from multiple expert models enables the training of ultra-compact student models that maintain competitive performance while reducing model size by approximately half compared to the baseline models. We summarize the contribution of this paper as follows:

\begin{itemize}[leftmargin=*]
    \item \textbf{Novelty:} To the best of our knowledge, this study is the first to apply the Mixture of Experts (MoE) framework to knowledge distillation for language models of code.
    

    \item \textbf{Technique:} We propose \textit{MoEKD}, a knowledge distillation framework that realizes a multi-teacher-to-student learning paradigm using a Mixture of Experts architecture to aggregate knowledge from multiple specialized expert code models and transfer it to a compact student model. \textit{MoEKD} enhances both predictive performance and adversarial robustness of compressed student models compared to state-of-the-art knowledge distillation approaches, as demonstrated through extensive empirical evaluation.
    
    \item \textbf{Open science:} we have open-sourced our replication package to facilitate future research\footnote{\url{https://doi.org/10.6084/m9.figshare.31144822}}.
\end{itemize}

\section{Preliminaries}
\label{back}


\subsection{Mixture of Experts (MoE)}
\label{MoEBack}
The Mixture of Experts framework is a collaborative learning approach that integrates multiple specialized models, known as experts, each dedicated to distinct regions of the input space. First introduced by Jacobs et al. \cite{jacobs1991adaptive} in 1991, this methodology partitions the input space into subspaces, trains individual experts within each partition, and aggregates their outputs via a routing mechanism, typically a gating function. Generally, a MoE architecture consists of three primary components. \textbf{Experts: } a set of specialized models, each trained to identify patterns within distinct regions of the input space or to address specific subtasks. While experts may share the same architecture, they learn different representations. Experts may be implemented as independent neural networks or specialized submodels that acquire complementary representations across different regions of the input space. \textbf{Gating Network (Router): } a routing mechanism that assigns each input to one or more experts by generating a set of weights or probabilities, which determine expert activation and their respective contributions to the final prediction. This router is typically implemented as a lightweight gating network that manages expert selection and contribution for each input instance. \textbf{Aggregation Mechanism (Combiner): } a component that integrates the outputs of the selected experts, typically using a weighted sum determined by the gating network’s outputs, to generate the final model prediction.




\begin{figure}[htbp]
\centering
\includegraphics[width=6cm, height = 4.75cm]{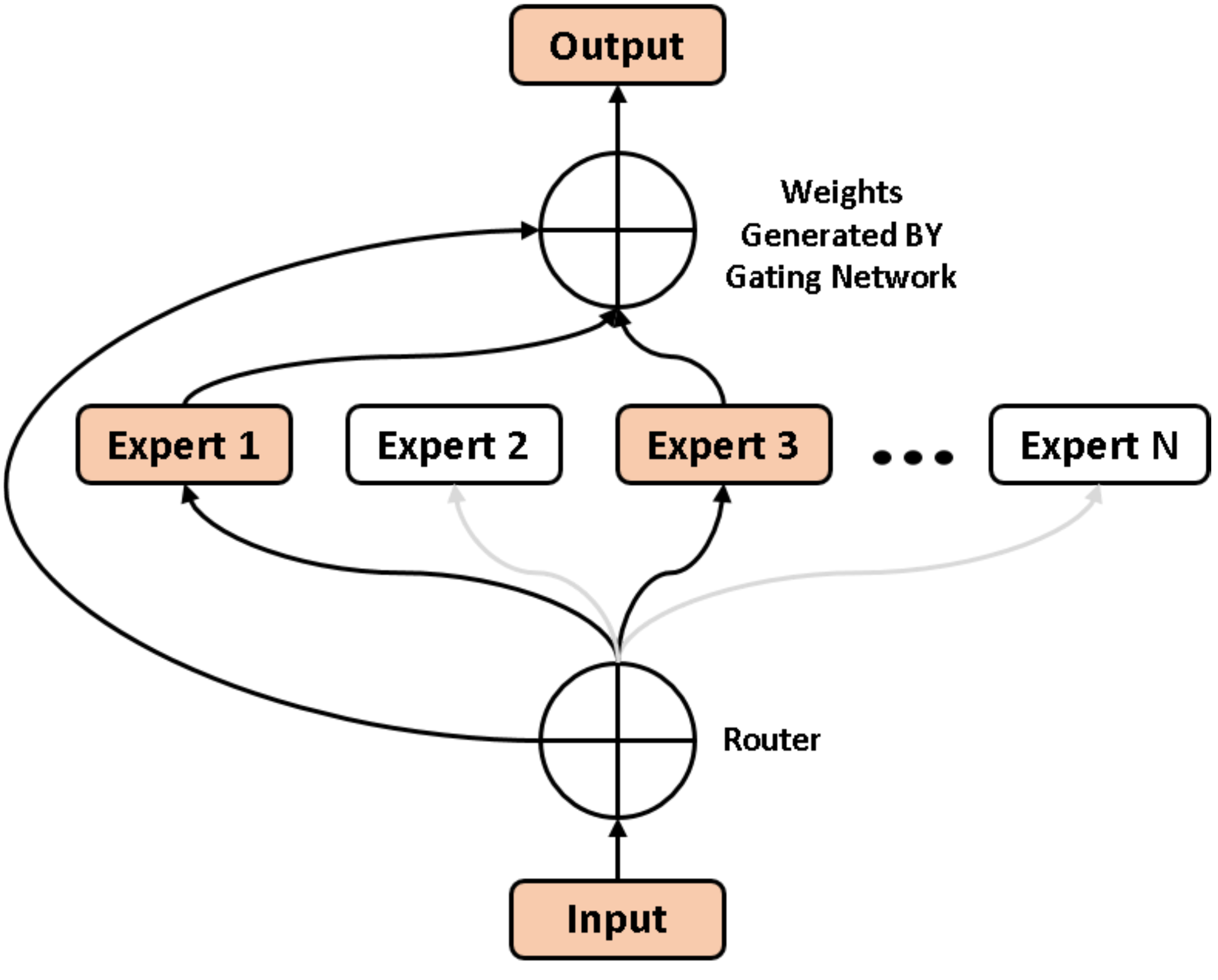}
\caption{An illustrative architecture of a MoE framework.} 
\label{MoEFramework}
\end{figure} 

Figure \ref{MoEFramework} presents a representative MoE architecture. Upon receiving an input, the routing module assesses the relevance of each expert and activates a subset for processing. In this scenario, the \textit{first} and \textit{third} experts are selected. The outputs from the selected experts are subsequently combined by the aggregation module to produce the final output. Training the MoE framework involves either jointly learning the routing parameters with the expert parameters or selectively optimizing a subset of these parameters depending on the deployment setting \cite{yuksel2012twenty}.

\subsection{Knowledge Distillation}
\label{KDBack}
Knowledge distillation (KD) \cite{hinton2015distilling} involves transferring knowledge from a large, expressive teacher model to a smaller, more efficient student model. Rather than relying exclusively on hard labels, the student is trained with teacher-provided soft knowledge, such as probability distributions or logits, often using unlabeled data. This approach enables the student to approximate the teacher’s functional behavior and decision boundaries, thereby preserving much of the teacher’s generalization ability while significantly reducing model size and computational requirements. KD methods can be categorized as either task-specific or task-agnostic, depending on how compressed models are applied \cite{xu2024surveyKD}. Task-specific distillation compresses a teacher model that has already been fine-tuned for a downstream task, allowing the student model to be directly deployed for the same task. In contrast, task-agnostic distillation compresses a pre-trained model prior to downstream adaptation, a process that typically requires expensive, large-scale training. Given the substantially higher computational and time costs associated with task-agnostic approaches, this study focuses on task-specific distillation, which provides a more practical and resource-efficient solution for software analytics tasks \cite{shi2022compressing}.

\section{Methodology}
\label{method}

\subsection{Problem Formulation}
\label{MoEAnalogy}
A vanilla knowledge distillation typically relies on a single teacher model to supervise the training of a compact student model. While this approach effectively transfers average task behavior, it may be limited if the knowledge obtained from the teacher does not adequately represent diverse or challenging decision regions, especially those sensitive to adversarial perturbations. For example, complex software engineering tasks such as vulnerability detection encompass a wide range of patterns, behaviors, and semantic cues that are unlikely to be uniformly captured by a single model \cite{yang2025one}. In these complex learning scenarios, robust understanding often emerges from exposure to multiple perspectives rather than instruction from a single source. This situation is analogous to higher education, where students do not depend on a single instructor to master an entire curriculum. Instead, learning is distributed among multiple teachers, each specializing in a specific subject area such as physics, mathematics, or chemistry. Such specialization enables each teacher to provide deeper, more focused knowledge within their domain, allowing students to develop a more comprehensive and robust understanding.


Extending this analogy, knowledge distillation can similarly benefit from supervision by multiple expert models (teachers) rather than a single teacher. In this context, each expert may specialize in capturing distinct aspects of the task, such as particular vulnerability categories, recurring code patterns, semantic structures, or features relevant to adversarial robustness. Aggregating knowledge from these specialized experts exposes the student model to a broader, more diverse set of learning features that may not be fully represented by a single teacher's knowledge. A \textbf{mixture of experts} framework provides a principled abstraction for this setting by enabling the systematic extraction and integration of expert knowledge during distillation. \textit{We hypothesize that distilling knowledge from multiple specialized experts can improve task performance and enhance robustness compared with single-teacher distillation.}

Building on the multi-expert learning paradigm, the knowledge distillation setting for this study is formalized as follows. Let $\mathcal{D} = \{(x_i, y_i)\}_{i=1}^{N}$ represent a labeled dataset for a software engineering task, where $x_i$ denotes a code fragment and $y_i$ its corresponding label. Consider a set of expert models $\{T_1, \ldots, T_K\}$, each specializing in distinct aspects of the task, and a compact student model $S$. The goal of KD is to train $S$ to effectively utilize the knowledge imparted by the expert models. Unlike conventional approaches that rely on a single teacher, this study investigates a Mixture of Experts scenario in which the student receives an aggregated knowledge, $\mathcal{A}(T_1, \ldots, T_K)$. Accordingly, the problem addressed in this study is to determine whether distilling knowledge from a mixture of experts improves the predictive performance and adversarial robustness of the student model relative to single-teacher distillation under both standard and adversarial evaluation settings.

To address this problem, we propose \textit{MoEKD}, a knowledge distillation framework that integrates a Mixture of Experts (MoE) mechanism to systematically extract and aggregate knowledge from multiple specialized expert (teacher) models. The subsequent section outlines the proposed framework, including the construction of expert models, the aggregation of expert outputs, and the training procedure for transferring the aggregated knowledge to the student model.

\subsection{Approach: \textit{MoEKD} Framework}
\label{MoEKDDesign}
\begin{figure*}[t]
\centering
\includegraphics[width=12cm, height = 6.5cm]{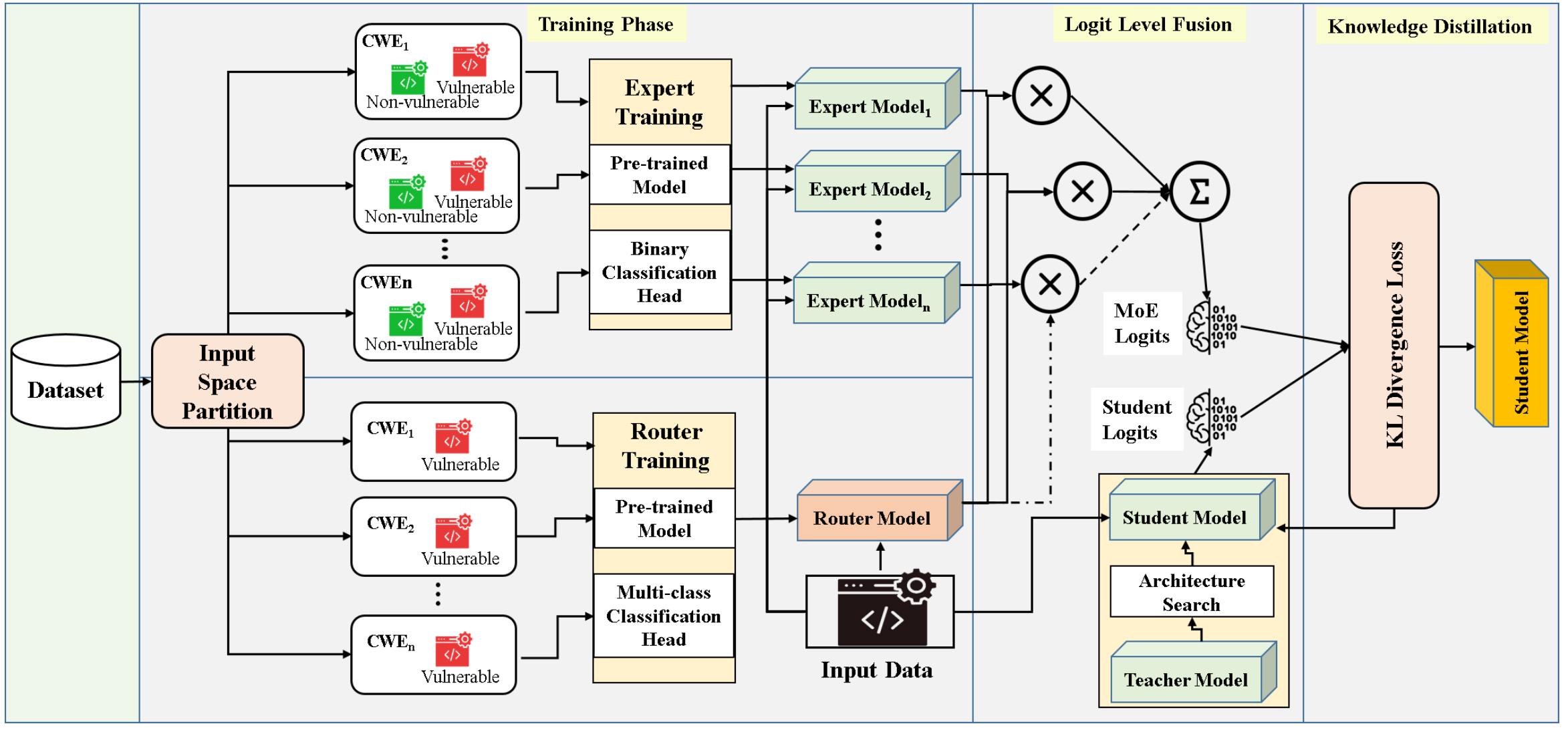}
\caption{Overview of the proposed \textit{MoEKD} framework, which trains multiple specialized experts and a router, aggregates selected expert outputs at the logit level, and distills the fused knowledge into a compact student model.} 
\label{MoEKDFramework}
\vspace{-10pt}
\end{figure*}

The proposed \textit{MoEKD} framework leverages a Mixture of Experts paradigm to extract specialized knowledge from multiple expert (teacher) models for knowledge distillation. Figure \ref{MoEKDFramework} illustrates the overall \textit{MoEKD} pipeline. The framework consists of three sequential phases: (i) expert and router training, where multiple specialized experts are learned from partitioned input spaces and a router is trained to estimate the relevance of each expert; (ii) output aggregation and logit level fusion, where the router dynamically selects the top-$k$ experts and combines their outputs into a unified knowledge; and (iii) knowledge distillation, where the aggregated expert knowledge is transferred to a compact student model using KL divergence-based supervision without hard labels. The following subsections describe the design of each component of the \textit{MoEKD} framework, using the BigVul \cite{fan2020ac} vulnerability detection dataset as an illustrative example.

\subsubsection{Training Phase}
\label{TrainMoEKD}
The training phase of \textit{MoEKD} consists of three key steps: partitioning the input space, training specialized expert models, and training the router to estimate the relevance of each expert. Following the study by Yang et al. \cite{yang2025one}, we train the expert and router models separately to reduce training complexity.

\textbf{Input space partitioning via CWE taxonomy: }An important design consideration in the MoE framework is the partitioning of the input space among experts, since effective specialization depends on a meaningful division of inputs. The goal is to assign each expert a subset of inputs with coherent, recurring patterns, allowing them to capture task-specific nuances more effectively than a generalist model. In vulnerability detection, this strategy aligns with categorization schemes that group common software system weaknesses, such as buffer overflows, improper authentication, and information leakage, based on shared characteristics. \cite{christey2013common}.

The Common Weakness Enumeration (CWE) taxonomy guides input space partitioning because distinct vulnerability categories exhibit unique code patterns and semantic properties that necessitate specialized modeling. Nevertheless, assigning an expert to each CWE type is impractical given the large number of categories, as demonstrated by the BigVul dataset \cite{fan2020ac}, which includes 88 CWE types and exhibits a highly imbalanced instance distribution. The limited representation of many CWE categories further complicates the development of reliable experts and increases training overhead. Consequently, following the methodology outlined by Yang et al. \cite{yang2025one}, the hierarchical structure of the CWE taxonomy is employed, and input splitting is performed at a coarse level by selecting top-level CWE categories. This method clusters related vulnerabilities with similar characteristics, enabling each expert to generalize across semantically related subcategories. CWE categories with insufficient training instances are consolidated into a single group to ensure stable expert training and to control model complexity. Furthermore, consistent with established knowledge distillation protocols, the training data is divided into two separate subsets: one for training the teacher model and the other for extracting soft knowledge to guide student training. To ensure reliable expert learning in this context, each CWE group is assigned at least 100 training samples, following the recommendation of Yang et al. \cite{yang2025one}. Accordingly, 10 input subspaces are constructed, each corresponding to a specific CWE category: \texttt{CWE-66}, \texttt{CWE-707}, \texttt{CWE-399}, \texttt{CWE-264}, \texttt{CWE-284}, \texttt{CWE-682}, \texttt{CWE-691}, \texttt{CWE-189}, \texttt{CWE-unknown}, and \texttt{CWE-other}. An expert model is trained for each subspace, specializing in a particular vulnerability.


\textbf{Expert training: }Each expert model is trained to specialize in a specific vulnerability subspace, aiming to extract focused and discriminative knowledge for subsequent transfer to the student model via knowledge distillation. For a given vulnerability category, the expert is optimized to distinguish code instances associated with that category from all other code, thereby enabling the model to learn category-specific patterns and decision boundaries. During training, code samples from the target vulnerability category are labeled as positive instances, whereas all other samples, including those from other categories and non-vulnerable code, are labeled as negative instances. This approach encourages each expert to capture fine-grained characteristics relevant to its assigned subspace.

For each vulnerability category $\mathrm{CWE}_i$, a dedicated expert model is trained to estimate the probability that a code fragment $c$ is vulnerable. Each expert is instantiated by fine-tuning a pre-trained code language model, such as CodeBERT, and is augmented with a binary classification head tailored to its assigned vulnerability subspace. The expert generates a prediction $P(\mathrm{vul} \mid c, \mathrm{CWE}_i, \theta_e^i)$, where $\theta_e^i$ represents the parameters of the $i$-th expert. Training is conducted on a category-specific dataset by minimizing a binary cross-entropy loss defined over vulnerable and non-vulnerable instances. This approach enables each expert to learn discriminative representations specific to its target vulnerability type.

\textbf{Router training: } The router is trained to identify which expert models are most relevant for a given input, enabling effective extraction of expert knowledge during distillation. Its role is to capture distinguishing characteristics of Common Weakness Enumeration (CWE) categories, ensuring that the student model receives guidance from the most suitable experts. Router training is therefore formulated as a multi-class classification task over CWE categories, with the goal of predicting the probability that an input code fragment belongs to each CWE-specific subspace. Given an input code fragment $c$, the objective is to assign it to one of the CWE categories $\mathrm{CWE}_i \in \{\mathrm{CWE}_1, \mathrm{CWE}_2, \ldots, \mathrm{CWE}_N\}$, where $N$ denotes the total number of CWE types and corresponds to the number of experts. To this end, the routing model is trained to estimate the probability that $c$ belongs to a specific CWE category $\mathrm{CWE}_i$, denoted as $g_i(\mathrm{CWE}_i \mid c, \theta_g)$, where $\theta_g$ represents the parameters of the router.

Similar to the expert models, the router is built on a pre-trained code language model (e.g., CodeBERT) that serves as the backbone and is augmented with a multi-class classification head to predict CWE categories. The distribution of CWE types is highly imbalanced and exhibits a pronounced long tail, a characteristic commonly observed in vulnerability datasets \cite{zhou2023devil}. To address this imbalance during router training, the focal loss function \cite{lin2017focal} is employed, as recommended by Yang et al. \cite{yang2025one}, instead of the standard cross-entropy loss. The focal loss prioritizes hard-to-classify examples and reduces the influence of dominant classes, thereby improving the router’s capacity to accurately model rare CWE categories. The overall focal loss for the router model over the training dataset $\mathcal{D}_R$ is defined as:

\begin{equation}
\mathcal{L}_R(\mathcal{D}_R, \theta_g) =
- \frac{1}{N_R}
\sum_{j=1}^{N_R}
\sum_{i=1}^{N}
\alpha_i \left(1 - \hat{t}_{ji}\right)^{\gamma}
t_{ji} \log\left(\hat{t}_{ji}\right),
\end{equation}

where $N$ denotes the number of CWE categories, $N_R$ is the number of training instances in the router training dataset $\mathcal{D}_R$, $\alpha_i$ is the class balancing factor for $\mathrm{CWE}_i$, $\gamma$ is the focusing parameter, $t_{ji}$ represents the ground truth label of code instance $c_j$ for class $\mathrm{CWE}_i$, and $\hat{t}_{ji} = g_i(\mathrm{CWE}_i \mid c_j, \theta_g)$ denotes the predicted probability assigned by the router that $c_j$ belongs to class $\mathrm{CWE}_i$.

\subsubsection{Expert Aggregation and Logit Level Fusion}
\label{MoEKDAggLogit}

We employ a Mixture of Experts framework in which a router model dynamically selects a subset of specialized expert models for each input instance and integrates their outputs at the logit level. Let $\mathcal{E} = \{e_1, e_2, \ldots, e_{N}\}$ denote a set of $N$ expert models, where each expert $e_j$ is trained as a binary classifier and produces two output logits corresponding to the non-vulnerable and vulnerable classes. Given an input code fragment $x_i$, the router model outputs a probability distribution over the experts, $\mathbf{p}_i = [p_{i,1}, p_{i,2}, \ldots, p_{i,N}], \quad \sum_{j=1}^{N} p_{i,j} = 1,$ where $p_{i,j} = g_j(x_i)$ denotes the router output probability indicating the relevance of expert $e_j$ for input $x_i$.

Rather than aggregating outputs from all experts, we select only the most relevant ones to balance expressiveness and computational efficiency. Specifically, for each input sample, we select the top-$k$ experts with the highest router probabilities. Let $\mathcal{K}_i = \{k_{i,1}, k_{i,2}, \ldots, k_{i,k}\}$ denote the indices of the selected experts such that $p_{i,k_{i,1}} \geq p_{i,k_{i,2}} \geq \cdots \geq p_{i,k_{i,k}} \geq p_{i,j}, \quad \forall j \notin \mathcal{K}_i.$


The router probabilities corresponding to the selected experts are renormalized to obtain convex combination weights, $w_{i,m} = \frac{p_{i,k_{i,m}}}{\sum_{m'=1}^{k} p_{i,k_{i,m'}}}, \quad m \in \{1, \ldots, k\},$ ensuring that $\sum_{m=1}^{k} w_{i,m} = 1$. These weights control the relative contribution of each selected expert during aggregation. Each selected expert $e_{k_{i,m}}$ produces a logit vector $\boldsymbol{\ell}_{i,k_{i,m}} \in \mathbb{R}^{2}$, representing unnormalized scores for the two output classes. The final expert aggregated representation for input $x_i$ is obtained by fusing the expert logits through a weighted summation,
\begin{equation}
\boldsymbol{\ell}_i^{\text{MoE}} = \sum_{m=1}^{k} w_{i,m} \cdot \boldsymbol{\ell}_{i,k_{i,m}} .
\end{equation}

Aggregation at the logit level preserves the relative confidence structure learned by each expert and avoids distortions associated with premature probability normalization. The resulting fused logits, $\boldsymbol{\ell}_i^{\text{MoE}}$, serve as the final supervisory knowledge used for downstream evaluation and knowledge distillation.

It is important to note that the router is trained exclusively on vulnerable code and, consequently, does not explicitly learn routing behavior for non-vulnerable inputs. This situation raises questions about how non-vulnerable samples are handled during logit fusion. In \textit{MoEKD}, non-vulnerable inputs are routed to the experts according to their similarity to vulnerable code patterns, and each expert is trained on both a specific CWE vulnerability and non-vulnerable code, which enables reliable processing of non-vulnerable inputs. Consequently, even if non-vulnerable code is routed to different experts, this does not pose a limitation, as all experts can handle non-vulnerable instances by design \cite{yang2025one}.

\subsubsection{Knowledge Distillation}
\label{KDMoEKD}

This Section outlines the knowledge distillation process in \textit{MoEKD}, including the selection of compact student architectures and the distillation objective for transferring aggregated expert knowledge to the student model.

\textbf{Student model architecture: }The student models in the proposed \textit{MoEKD} framework reuse compact architectures derived from existing model compression techniques rather than introducing new architectural designs. The selection of student architectures is guided by two key constraints: \textit{deployment feasibility} and \textit{experimental control}. First, we target compact models suitable for integration into integrated development environments and therefore impose a strict model size constraint of approximately 3 MB. This choice builds on prior work on deployable vulnerability-detection models and aligns with practical usage scenarios in software development workflows \cite{svyatkovskiy2021fast, aye2020sequence}. Second, robustness-aware objectives are intentionally excluded from the original multi-objective compression process during the identification of candidate student architectures. This strategy isolates the effect of Mixture of Experts supervision and enables assessment of whether improvements in predictive performance and adversarial robustness can be achieved solely through aggregated expert knowledge distillation, without reliance on robustness-oriented architectural optimization.

Under these constraints, we select student architectures generated by \textit{Compressor} and \textit{AVATAR}, two established compression frameworks that produce reduced capacity variants of large pretrained code language models via structured compression strategies. Among the Pareto-optimal configurations produced by these methods, we select those that satisfy the imposed size constraint while maintaining sufficient model capacity. The selected architectures remain fixed throughout training, ensuring that any observed performance or robustness gains can be attributed to the proposed \textit{MoEKD} framework rather than differences in model structure.

\textbf{Student training objective: } The student training stage in \textit{MoEKD} aims to facilitate effective learning in a compact student model by leveraging aggregated knowledge from multiple specialized experts. For each input code fragment $x_i$, the expert aggregation module generates fused logits $\boldsymbol{\ell}_i^{\text{MoE}}$, which represent complementary knowledge from the most relevant experts as determined by the router. These fused logits provide soft supervision to guide the student model during knowledge distillation.

Let $\boldsymbol{\ell}_i^{S}$ denote the output logits produced by the student model for input $x_i$. Rather than combining hard-label supervision with distillation, we formulate student training as a knowledge distillation problem, where the objective is to align the student’s predictive distribution with that of the aggregated experts. This design choice enables the student model to capture fine-grained decision boundaries and inter-class relationships learned by the expert ensemble, without being constrained by hard labels. Formally, the student model is trained by minimizing the Kullback-Leibler \cite{kullback1951information} divergence between the softened output distributions of the aggregated experts and the student model.

\begin{equation}
\mathcal{L}_{S} =
-\frac{1}{n}
\sum_{i=1}^{n}
\mathrm{softmax}\!\left(\frac{\boldsymbol{\ell}_i^{\text{MoE}}}{T}\right)
\log
\left(
\mathrm{softmax}\!\left(\frac{\boldsymbol{\ell}_i^{S}}{T}\right)
\right)
T^{2}
\label{KLDLoss}
\end{equation}

In Eq.~\eqref{KLDLoss}, $\boldsymbol{\ell}_i^{\text{MoE}}$ and $\boldsymbol{\ell}_i^{S}$ denote the aggregated expert logits and the student logits for input $x_i$, respectively, and $T$ is the temperature parameter controlling the smoothness of the output distributions. The aggregated expert logits $\boldsymbol{\ell}_i^{\text{MoE}}$ remain fixed during training, while the student logits $\boldsymbol{\ell}_i^{S}$ are updated to minimize the loss.

\section{Empirical Evaluation}
\label{MoEKDEval}
Following the description of the proposed \textit{MoEKD} framework and its components, we present an empirical evaluation to assess its effectiveness in enhancing predictive performance and robustness of compressed student models relative to SOTA KD approaches. This evaluation addresses the following research questions:

\begin{itemize}[left=0pt]
    \item \textbf{RQ1}: \textit{Does a mixture of experts-based knowledge distillation enhance the performance of student models relative to single-teacher distillation?}
    \item \textbf{RQ2}: \textit{Does a mixture of experts-based knowledge distillation improve the adversarial robustness of student models relative to single-teacher distillation?}
    \item \textbf{RQ3}: \textit{Does \textit{MoEKD} retain its performance advantages over single-teacher distillation under more aggressive model compression?}
\end{itemize}

\subsection{Knowledge Distillation Baselines}
\label{KDBase}
Several knowledge distillation approaches have been proposed for language models of code, including \textit{Compressor} \cite{shi2022compressing}, \textit{AVATAR} \cite{shi2024greening}, \textit{MORPH} \cite{panichella2025metamorphic}, \textit{PIONEER} \cite{liu2026pioneer}, and \textit{SODA} \cite{chen2025smaller}. In this study, we restrict our baselines to methods that produce compact student models of size comparable to 3 MB. Except for \textit{SODA}, all of these approaches generate student models of approximately 3 MB in size. Both \textit{MORPH} and \textit{PIONEER} explicitly incorporate robustness-oriented objectives during knowledge distillation. As our study intentionally excludes robustness-aware optimization to isolate the effect of Mixture of Experts supervision, these methods are not considered as baselines. Based on these criteria, we select \textit{Compressor} and \textit{AVATAR} as baseline approaches for comparison.

\subsection{Datasets and Subjects}
\label{datasets}
To ensure a fair assessment of the performance and robustness improvements introduced by the mixture of expert-based knowledge distillation, we use the same models and tasks as those used by the baseline knowledge distillation–based approaches described in Section \ref{KDBase}. For the vulnerability detection task, the BigVul dataset \cite{fan2020ac} is used, as it provides CWE-type annotations necessary for input-space partitioning, as described in Section \ref{TrainMoEKD}. BigVul comprises vulnerable and non-vulnerable functions collected from more than 300 open-source C and C++ projects, spanning 88 CWE types.

To address the severe class imbalance in the original dataset, a balanced dataset is constructed using a structured preprocessing pipeline. All vulnerable functions are retained, while non-vulnerable functions are sampled in a project-stratified manner to preserve project-specific context. To further control for confounding effects related to function size, non-vulnerable samples are selected within a \textbf{20\%} range above or below the corresponding vulnerable function’s length, measured in lines of code. If no suitable match is available, a non-vulnerable function is randomly selected from the same project. The resulting dataset contains equal numbers of vulnerable and non-vulnerable functions and is randomly shuffled using a fixed seed to ensure reproducibility. In accordance with prior studies \cite{yang2025one}, the dataset is divided into 80\% for training, 10\% for validation, and 10\% for testing. Consistent with the standard single-teacher knowledge distillation approach, the training split is further divided into two equal subsets: one for training the experts and router, and the other for generating soft knowledge to train the student model. This preprocessing produces the final dataset used to evaluate the \textit{MoEKD} framework.

CodeBERT \cite{feng2020codebert} and GraphCodeBERT \cite{guo2020graphcodebert} serve as the underlying language models for code in this study. CodeBERT is a pretrained transformer-based \cite{vaswani2017attention} model that adopts the architecture of RoBERTa \cite{liu2019roberta}, comprising 12 transformer layers with multi-head self-attention and a hidden dimensionality of 768. During pretraining, CodeBERT is optimized with two objectives: masked language modeling, which learns contextual token representations by predicting masked tokens, and replaced token detection, which distinguishes original tokens from substituted ones. GraphCodeBERT extends CodeBERT by incorporating structural information derived from program data-flow graphs. While maintaining the same transformer architecture, GraphCodeBERT integrates data-flow relationships between code tokens during pretraining and introduces an additional objective to predict the existence of data-flow edges between nodes.

\subsection{Evaluation Metrics}
\label{EvalMoEKD}
The effectiveness of the proposed \textit{MoEKD} framework is evaluated in two complementary settings: accuracy before adversarial attacks and robustness after adversarial attacks. This two-stage evaluation assesses whether \textit{MoEKD} improves both predictive performance and robustness relative to single-teacher distillation.

\textbf{Pre-attack performance}: Predictive performance under standard conditions is evaluated by measuring classification \textit{accuracy} on the clean test set. Accuracy is the primary metric because the selected knowledge distillation baselines consistently use it to compare student models with their corresponding teacher models. This approach ensures a fair and direct comparison with prior work and enables assessment of whether the Mixture of Experts-based distillation preserves or enhances the predictive performance of compact student models.

\textbf{Post-attack robustness}: Robustness is assessed by evaluating student models in adversarial settings using three state-of-the-art identifier renaming-based attack strategies: \textbf{ALERT} \cite{yang2022natural}, \textbf{Metropolis Hastings Modifier (MHM)} \cite{zhang2020generating}, and \textbf{WIR-Random} \cite{zeng2022extensive}. ALERT applies greedy and evolutionary search strategies over both operational and natural code semantics, using 30 candidate substitutions per identifier. MHM performs iterative identifier substitution via Metropolis–Hastings sampling, with a maximum of 100 iterations and 30 candidate evaluations per step. WIR-Random ranks identifiers according to their influence on model predictions and applies sequential random substitutions. These attacks maintain program semantics while renaming identifiers to induce misclassification. Robustness is quantified using the Attack Success Rate (\%ASR), which measures the proportion of originally correctly classified samples that are misclassified following the attack. A lower \%ASR reflects greater robustness. In all cases, adversarial attacks are applied only to samples that are correctly classified prior to perturbation, as recommended by the previous research \cite{yang2022natural, zhang2020generating}.

\subsection{Experimental Setting}
\label{ExpSetting}
All experiments were conducted on a Linux server running Ubuntu 22.04.5 LTS, equipped with an Intel Xeon Platinum 8356H CPU, 3.0 TB RAM, and three NVIDIA A100 GPUs with 80 GB memory each. Random seeds were fixed across all experiments to ensure reproducibility. CodeBERT was fine-tuned following the training pipeline and hyperparameter configurations defined in the CodeXGLUE benchmark \cite{lu2021codexglue}, while GraphCodeBERT was trained using the hyperparameter settings reported in its original work \cite{guo2020graphcodebert}. CodeGraph and GraphCodeBERT were used as the backbone of the experts and router model, similar to single-teacher distillation, to mitigate experimental bias and ensure a fair comparison.

The student models utilize compact architectures generated by the baseline compression techniques \textit{Compressor} and \textit{AVATAR}. The exact student architectures and training configurations identified by these baselines are reused without modification. This approach ensures that any observed differences in predictive performance and adversarial robustness can be attributed exclusively to the proposed Mixture of Experts-based knowledge distillation strategy, rather than to architectural changes or robustness-aware optimization. The architectural details of the compact student models are presented in Table \ref{STConfig}. Since the number of selected experts plays a crucial role in effective knowledge aggregation and directly affects the trade-off between inference time and performance, we follow the empirically validated top-$k$ expert selection setting reported by Yang et al. \cite{yang2025one}. Accordingly, we set $k = 2$ for all experiments.

\begin{table}[htbp]
\centering
\caption{Hyperparameter configurations of student model.}
\label{tab:compressor_avatar_config}
\resizebox{\columnwidth}{!}{%
\begin{tabular}{m{0.19\linewidth} | m{0.81\linewidth}}
\hline
\textbf{\textit{Compressor}} &
attention\_heads: 8, hidden\_dim: 96, intermediate\_size: 64, n\_layers: 12, vocab\_size: 1{,}000
\\
\hline
\textbf{\textit{AVATAR}} &
tokenizer: BPE, vocab\_size: 19{,}302, num\_hidden\_layers: 7, hidden\_size: 18, hidden\_act: GELU\_new, hidden\_dropout\_prob: 0.5, intermediate\_size: 1{,}491, num\_attention\_heads: 6, attention\_probs\_dropout\_prob: 0.4, max\_sequence\_length: 351, position\_embedding\_type: absolute, learning\_rate: 1e--3, batch\_size: 64
\\
\hline
\end{tabular}%
}
\vspace{-10pt}
\label{STConfig}
\end{table}

\section{Evaluation Results}
\label{EvalRes}

\subsection{\textit{RQ1: Pre-attack Performance}}
\label{RQ1EvalRes}

Table \ref{RQ1AccTab} presents the classification accuracy of uncompressed teacher models and their corresponding compressed student variants under standard (pre-attack) evaluation settings. As expected, the uncompressed teacher models achieve the highest accuracy, with CodeBERT and GraphCodeBERT attaining scores of 0.65 and 0.64, respectively. 

Model compression inevitably involves trade-offs, as reducing the capacity of a code language model can degrade predictive performance, including lower accuracy on downstream tasks \cite{shi2022compressing}. In particular, compressed student models obtained via single-teacher distillation exhibit a noticeable reduction in predictive performance. For example, the \textit{Compressor}-based student model achieves accuracies of 0.54 (CodeBERT) and 0.59 (GraphCodeBERT), while the \textit{AVATAR}-based student model attains 0.57 and 0.58, respectively. This degradation underscores the challenge of maintaining predictive performance when compressing language models of code using conventional single-teacher supervision.

\begin{table}[htbp]
\centering
\caption{Accuracy comparison of teacher models and compressed student variants.}
\label{RQ1AccTab}
\begin{threeparttable}[t]
\begin{tabular}{l l cc}
\hline
\textbf{Method}                                                                         & \textbf{Model}   & \textbf{CB}   & \textbf{GCB}  \\ \hline
Uncompressed                                                                   & Teacher & 0.65 & 0.64 \\ \hline
\rowcolor[HTML]{EFEFEF} 
\cellcolor[HTML]{EFEFEF}                                                       & MoE\_CP & 0.61 & 0.62 \\ \cline{2-4} 
\rowcolor[HTML]{EFEFEF} 
\multirow{-2}{*}{\cellcolor[HTML]{EFEFEF}\textit{Compressor}} & CP      & 0.54 & 0.59 \\ \hline
\rowcolor[HTML]{ECF4FF} 
\cellcolor[HTML]{ECF4FF}                                                       & MoE\_AV & 0.63 & 0.62 \\ \cline{2-4} 
\rowcolor[HTML]{ECF4FF} 
\multirow{-2}{*}{\cellcolor[HTML]{ECF4FF}\textit{ AVATAR }}   & AV      & 0.57 & 0.58 \\ \hline
\end{tabular}

\begin{tablenotes} [flushleft] \tiny
  \item[] \parbox[t]{\linewidth}{
    \item[] CB = CodeBERT; GCB = GraphCodeBERT; CP = Compressor; AV = AVATAR; MoE\_CP and MoE\_AV denote \textit{MoEKD}-based students.
  }
\end{tablenotes}

\end{threeparttable}

\vspace{-10pt}
\end{table}

In contrast, student models trained using the proposed \textit{MoEKD} framework consistently outperform their single-teacher counterparts across both backbone models. Specifically, the MoE-enhanced \textit{Compressor} student (MoE\_CP) improves accuracy from 0.54 to 0.61 for CodeBERT, representing a relative improvement of approximately 13.0\%, and from 0.59 to 0.62 for GraphCodeBERT, yielding an improvement of about 5.1\%. Similarly, for \textit{AVATAR}-based students, \textit{MoEKD} improves accuracy from 0.57 to 0.63 for CodeBERT, representing a relative gain of approximately 10.5\%, and from 0.58 to 0.62 for GraphCodeBERT, corresponding to an improvement of about 6.9\%. These results demonstrate that distilling knowledge from a mixture of specialized experts yields significantly stronger predictive performance than single-teacher distillation, even under identical student architectures and training configurations. 

\begin{center}
\begin{tcolorbox}[
    enhanced,
    attach boxed title to top left={yshift=-3mm,yshifttext=-1mm}, 
    colback=mycolor_box,                 
    colframe=black,                
    colbacktitle= mycolor_title,            
    coltitle=black,                
    title=Answer to RQ1,            
    fonttitle=\bfseries,           
    boxed title style={size=small},
    width=0.49\textwidth,            
    boxsep=1mm,                    
    left=0mm,                      
    right=0mm, 
    bottom=0mm,
]

\textit{MoEKD} consistently improves the predictive accuracy of compressed student models over single-teacher distillation for both CodeBERT and GraphCodeBERT in the vulnerability detection task, achieving relative performance gains ranging from approximately \textbf{5.1\% to 13.0\%} under identical knowledge distillation settings.

\end{tcolorbox}
\label{RQ1Result}
\end{center}


\subsection{\textit{RQ2: Post-attack Robustness}}
\label{RQ2EvalRes}
\begin{figure}[htbp]
  \centering
  \subfigure[CodeBERT + \textit{Compressor}]{\label{CBCP}\includegraphics[width=4cm, height=3cm]{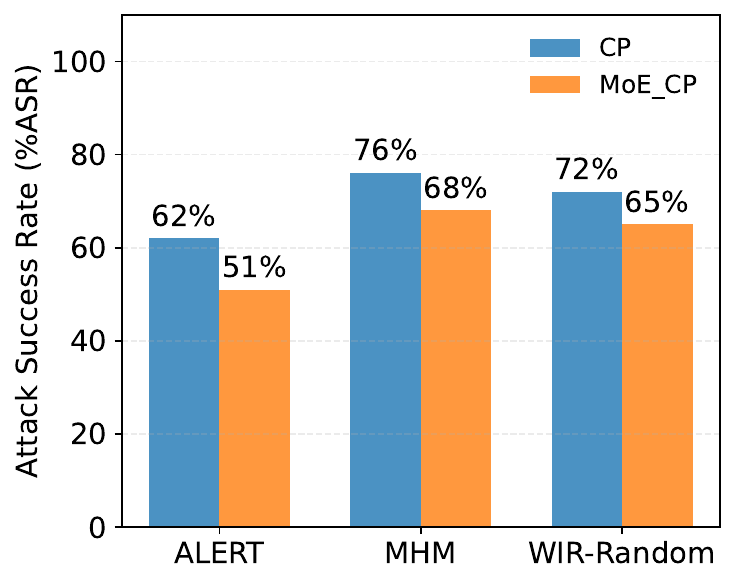}}
  \subfigure[CodeBERT + \textit{AVATAR}]{\label{CBAV}\includegraphics[width=4cm, height=3cm]{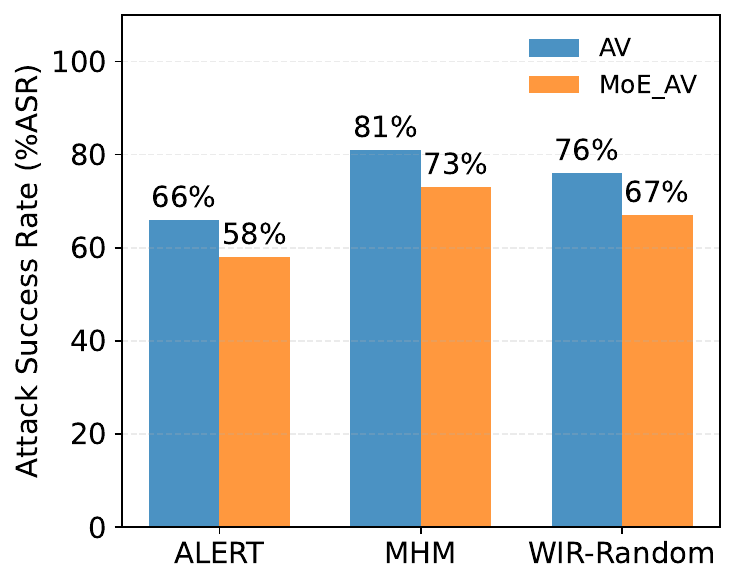}}
  \subfigure[GraphCodeBERT + \textit{Compressor}]{\label{GCBCP}\includegraphics[width=4cm, height=3cm]{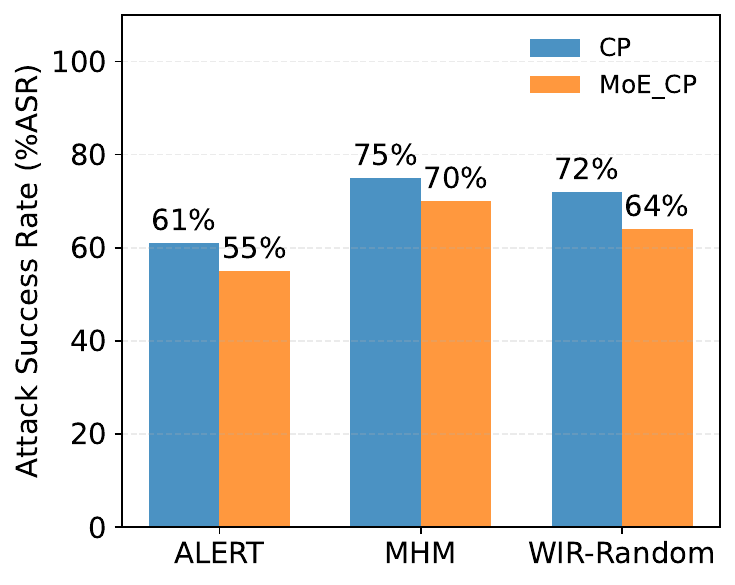}}
  \subfigure[GraphCodeBERT + \textit{AVATAR}]{\label{GCBAV}\includegraphics[width=4cm, height=3cm]{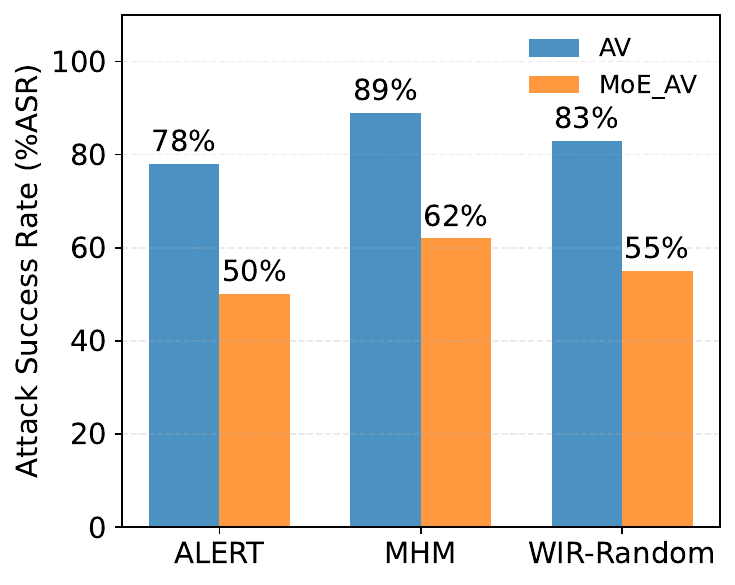}}
  \caption{Adversarial robustness comparison of single-teacher and \textit{MoEKD}-based compressed student models across ALERT, MHM, and WIR-Random attacks, measured using the Attack Success Rate (\%ASR).}
  \label{RQ2Robustness}
  \vspace{-10pt}
\end{figure}

Having established the predictive performance gains achieved by compressed student models under the \textit{MoEKD} framework, we now investigate its effectiveness in enhancing these models' adversarial robustness relative to single-teacher distillation. Prior work by Panichella et al. \cite{panichella2025metamorphic} has shown that knowledge-distilled student models typically exhibit lower adversarial robustness than their teacher counterparts; therefore, our robustness analysis compares \textit{MoEKD}-distilled students with those trained using single-teacher distillation. Figure \ref{RQ2Robustness} reports the Attack Success Rate (\%ASR) under three adversarial attacks for compressed models obtained from \textit{MoEKD}, \textit{Compressor}, and \textit{AVATAR}.

Since lower \%ASR indicates stronger robustness, across all evaluated settings, \textit{MoEKD}-distilled student models consistently exhibit improved adversarial robustness compared with their single-teacher distilled counterparts. For \textit{Compressor} architecture-based students, models distilled with \textit{MoEKD} consistently achieve greater adversarial robustness than those distilled with a single teacher across all three attack types. This is shown by a consistent reduction in \%ASR for both CodeBERT and GraphCodeBERT backbones. Across ALERT, MHM, and WIR-Random attacks, \textit{MoEKD}-distilled students show robustness improvements of approximately \textbf{6.7\%–17.7\%} compared to single-teacher distillation, depending on the attacks and backbone models.

A similar trend is observed among \textit{AVATAR} architecture-based students. \textit{MoEKD}-distilled students achieve robustness improvements of approximately \textbf{9.9\%–35.8\%} compared to single-teacher counterparts across all attacks. These gains are consistent for both CodeBERT and GraphCodeBERT models, showing that \textit{MoEKD}'s robustness benefits are not limited to a specific architecture or compression method. These results indicate that distilling knowledge from multiple specialized experts enables compact student models to better learn decision boundaries and retain robustness relevant information that cannot be adequately transferred by a single teacher, demonstrating the effectiveness of the \textit{MoEKD} framework.

To further determine whether the observed improvements in adversarial robustness are statistically significant, we conduct a paired statistical analysis comparing \textit{MoEKD}-distilled students with single-teacher counterparts across all attack settings. The Wilcoxon signed-rank \cite{woolson2007wilcoxon} test indicates that the robustness improvements achieved by \textit{MoEKD} are statistically significant, with a $p$-value of 0.006, suggesting that the observed gains are unlikely to be due to random variation. The corresponding Cliff’s delta \cite{cliff1993dominance} value of $\delta = 0.618$ indicates a large effect size, confirming that the improvements are both statistically and practically significant. Overall, these results provide strong empirical evidence that distilling knowledge from multiple specialized experts significantly enhances the adversarial robustness of compressed student models compared to single-teacher distillation.

\begin{center}
\begin{tcolorbox}[
    enhanced,
    attach boxed title to top left={yshift=-3mm,yshifttext=-1mm}, 
    colback=mycolor_box,                 
    colframe=black,                
    colbacktitle= mycolor_title,            
    coltitle=black,                
    title=Answer to RQ2,            
    fonttitle=\bfseries,           
    boxed title style={size=small},
    width=0.49\textwidth,            
    boxsep=1mm,                    
    left=0mm,                      
    right=0mm, 
    bottom=0mm,
]

\textit{MoEKD} consistently improves the adversarial robustness of compressed student models over single-teacher distillation for both CodeBERT and GraphCodeBERT across all evaluated attacks, achieving robustness gains ranging from approximately \textbf{6.7\% to 35.8\%} under identical distillation settings.

\end{tcolorbox}
\label{RQ2Result}
\end{center}

\subsection{\textbf{RQ3: Ablation Study on Knowledge Quality}}
While RQ1 and RQ2 show that distilling knowledge from the mixture of specialized experts improves predictive performance and adversarial robustness, it is unclear whether these gains stem from higher-quality aggregated knowledge. To assess the knowledge quality, we conduct an ablation study using ultra-compact CodeBERT-based student models of approximately 2 MB and 1.6 MB. The architectures of these models are selected from the Pareto-optimal configurations produced by \textit{Compressor} and \textit{AVATAR}.

\textbf{Accuracy under reduced model capacity: }
Table \ref{RQ3AccTab} shows that \textit{MoEKD} maintains its performance gains even with aggressive model compression, demonstrating the strength of aggregated expert knowledge. For both \textit{Compressor} and \textit{AVATAR} architectures, \textit{MoEKD} produces student models of 1.6 MB and 2 MB that consistently outperform the 3 MB single-teacher baselines. In the \textit{Compressor} architecture, \textit{MoEKD} increases accuracy from 0.54 to 0.59 and 0.60 at 1.6 MB and 2 MB, respectively, despite the smaller model size. Similarly, for \textit{AVATAR}-based students, \textit{MoEKD} achieves accuracies of 0.60 and 0.61 at 1.6 MB and 2 MB, exceeding the 3 MB single-teacher baseline accuracy of 0.57.

\begin{table}[htbp]
\centering
\caption{Accuracy comparison of \textit{MoEKD} and single-teacher distillation under varying model sizes.}
\label{RQ3AccTab}
\begin{tabular}{l l cc}
\hline
\multicolumn{1}{c}{\textbf{Architecture}} & \multicolumn{1}{c}{\textbf{Method}}     & \multicolumn{1}{c}{\textbf{Size}}           & \multicolumn{1}{c}{\textbf{Accuracy}} \\ \hline
                                            & \cellcolor[HTML]{EFEFEF}\textbf{CP}      & \cellcolor[HTML]{EFEFEF}\textbf{3 MB}        & \cellcolor[HTML]{EFEFEF}\textbf{0.54}  \\ \cline{2-4} 
                                            & \cellcolor[HTML]{EFEFEF}\textbf{MoE\_CP} & \cellcolor[HTML]{EFEFEF}\textbf{$\sim$1.6MB} & \cellcolor[HTML]{EFEFEF}\textbf{0.59}  \\ \cline{2-4} 
                                            & MoE\_CP                                  & $\sim$2MB                                    & 0.6                                    \\ \cline{2-4} 
\multirow{-4}{*}{\textit{Compressor}}                & MoE\_CP                                  & 3MB                                          & 0.62                                   \\ \hline
                                            & \cellcolor[HTML]{DAE8FC}\textbf{AV}      & \cellcolor[HTML]{DAE8FC}\textbf{3 MB}        & \cellcolor[HTML]{DAE8FC}\textbf{0.57}  \\ \cline{2-4} 
                                            & \cellcolor[HTML]{DAE8FC}\textbf{MoE\_AV} & \cellcolor[HTML]{DAE8FC}\textbf{$\sim$1.6MB} & \cellcolor[HTML]{DAE8FC}\textbf{0.60}  \\ \cline{2-4} 
                                            & MoE\_AV                                  & $\sim$2MB                                    & 0.61                                   \\ \cline{2-4} 
\multirow{-4}{*}{\textit{AVATAR}}                    & MoE\_AV                                  & 3MB                                          & 0.63                                   \\ \hline
\end{tabular}
\vspace{-8pt}
\end{table}

\textbf{Robustness under reduced model capacity: }
Table~\ref{RQ3RobustTab} reports adversarial robustness comparison, where lower \%ASR values indicate greater robustness. MoE\_CP generally outperforms the single-teacher baseline in most attack scenarios. With aggressive compression, the ~1.6 MB MoE\_CP model reduces \%ASR on ALERT from 0.62 to 0.59 and on Wir-Random from 0.72 to 0.70. For the MHM attack, the single-teacher baseline shows slightly better robustness than the ~1.6 MB MoE\_CP model (0.76 vs. 0.77). However, the ~2 MB MoE\_CP model improves robustness across all three attacks, achieving lower \%ASR than the baseline on ALERT (0.57 vs. 0.62), MHM (0.72 vs. 0.76), and Wir-Random (0.69 vs. 0.72). Overall, these results provide empirical evidence that aggregating knowledge from multiple specialized experts yields higher-quality knowledge, which, when transferred to the student model, enhances both the predictive performance and adversarial robustness of compressed student models compared to single-teacher distillation.

\begin{table}[htbp]
\centering
\caption{Adversarial robustness comparison of \textit{MoEKD} and single-teacher distillation under varying model sizes.}
\label{RQ3RobustTab}
\begin{tabular}{l l ccc}
\hline
\textbf{Method}  & \textbf{Size}        & \textbf{ALERT} & \textbf{MHM}                         & \textbf{Wir-Random} \\ \hline
\rowcolor[HTML]{DAE8FC} 
\textbf{CP}      & \textbf{3 MB}        & \textbf{0.62}  & \textbf{0.76}                        & \textbf{0.72}       \\ \hline
\rowcolor[HTML]{DAE8FC} 
\textbf{MoE\_CP} & \textbf{$\sim$1.6MB} & \textbf{0.59}  & {\color[HTML]{FE0000} \textbf{0.77}} & \textbf{0.7}        \\ \hline
MoE\_CP          & $\sim$2MB            & 0.57           & 0.72                                 & 0.69                \\ \hline
MoE\_CP          & 3MB                  & 0.51           & 0.68                                 & 0.65                \\ \hline
\end{tabular}
\end{table}


\begin{center}
\begin{tcolorbox}[
    enhanced,
    attach boxed title to top left={yshift=-3mm,yshifttext=-1mm}, 
    colback=mycolor_box,                 
    colframe=black,                
    colbacktitle= mycolor_title,            
    coltitle=black,                
    title=Answer to RQ3,            
    fonttitle=\bfseries,           
    boxed title style={size=small},
    width=0.49\textwidth,            
    boxsep=1mm,                    
    left=0mm,                      
    right=0mm, 
    bottom=0mm,
]

The performance and robustness gains achieved by \textit{MoEKD} under aggressive compression are driven by the quality of knowledge extracted from the mixture of specialized experts, demonstrating the effectiveness of Mixture of Experts-based knowledge distillation.


\end{tcolorbox}
\label{RQ3Result}
\end{center}

\section{Discussion}
\label{disc}
\subsection{Inference Cost and Training Overhead}
\label{DiscTraingCost}
When adopting \textit{MoEKD}, it is important to consider the increased computational overhead during training. \textit{MoEKD} uses multiple specialized expert models and a router, which increases training costs (e.g., energy consumption and carbon footprint) compared to single-teacher distillation. In our case, this overhead scales with the number of experts, leading to a roughly tenfold increase in teacher-side computation. Furthermore, \textit{MoEKD} incurs additional computational overhead during knowledge extraction. In contrast to single-teacher distillation, which queries only one teacher model, \textit{MoEKD} processes each input through the router and the top-$k$ selected experts (with $k=2$ in this study) to generate aggregated knowledge. Although \textit{MoEKD} increases teacher-side computation, the overhead is limited to the offline training phase and does not impact inference latency. At the inference phase, \textit{MoEKD} employs the same compact student architecture as baseline models, resulting in identical runtime cost, memory usage, and latency. By improving predictive performance and adversarial robustness without increasing inference cost, \textit{MoEKD} offers a favorable trade-off that prioritizes deployment efficiency and robustness, at the expense of higher one-time offline training cost.


\subsection{Router Limitation and Potential Future Direction}
\label{DiscRouterSelect}
The proposed \textit{MoEKD} framework uses a vanilla Mixture of Experts approach proposed by Jacobs et al. \cite{jacobs1991adaptive}, where knowledge extraction depends on effective input partitioning and appropriate expert selection. The router model controls both factors, making its performance critical to the framework's success. In our experiments, the trained router achieved about \textbf{62\%} accuracy, meaning many vulnerable code instances were not routed to the most suitable experts. While \textit{MoEKD} demonstrates measurable gains in performance and robustness compared to the baselines, there remains significant room for further improvement. In addition, \textit{MoEKD} currently relies on the vanilla Mixture of Experts formulation, and incorporating more advanced MoE architectures \cite{fedus2022switch, lepikhin2020gshard} may further improve expert knowledge aggregation and enable more effective distillation.




\section{Threats to Validity}
\label{threat}

\subsection{Internal Validity}
\label{internal}
A primary threat to internal validity is that observed performance differences could result from inconsistencies in fine-tuning strategies or data processing choices, rather than from the distillation methods themselves. To mitigate this threat, uniform training protocols, consistent hyperparameter settings, and identical data partitions were applied across all teacher and student models, in accordance with prior distillation studies \cite{shi2022compressing, shi2024greening, panichella2025metamorphic}. Additionally, experimental results may be affected by implementation errors in the training pipelines for teacher and student models. To minimize this threat, standardized training procedures from the \textit{CodeXGLUE} benchmark were adopted for teacher model fine-tuning and evaluation, and student baselines for the BigVul dataset were reproduced using the official replication packages of \textit{Compressor} and \textit{AVATAR}.

\subsection{Construct Validity}
\label{construct}
This study hypothesizes that aggregating knowledge from multiple expert models through a Mixture of Experts (MoE) mechanism facilitates more effective knowledge distillation than single-teacher supervision. Nevertheless, this assumption may not account for all factors influencing distillation outcomes, as the current analysis is restricted to sparse MoE configurations. Alternative architectures or routing strategies could potentially enable more effective knowledge aggregation. Additionally, observed performance improvements may also arise from architectural diversity rather than improved knowledge transfer alone. To mitigate this risk, we compare the proposed approach with single-teacher distillation baselines under identical architectures, training, and evaluation settings, while leaving a systematic analysis of alternative expert aggregation strategies to future work. Selecting adversarial attacks may limit construct validity, as classical attacks do not capture the full spectrum of adversarial strategies. However, these attacks introduce minimal perturbations while effectively targeting compressed models, suggesting that limited robustness under such settings likely indicates greater vulnerability to more advanced attacks.

\subsection{External Validity}
\label{external}
Threats to external validity concern the extent to which the findings of this study can be generalized beyond the specific models and tasks considered. The evaluation is conducted on the BigVul dataset using CodeBERT and GraphCodeBERT, both widely adopted for vulnerability detection. Although the results may not directly generalize to other datasets or model architectures, the proposed approach is model-agnostic and applicable to a broad range of code language models. Future research should examine the approach's applicability across additional datasets and model architectures to further assess its generalizability. To facilitate reproducibility and future research, comprehensive documentation of the methodology and experimental setup is provided, and all code and scripts used in the experiments are publicly released.

\section{Related Work}
\label{RW}
Model compression has been widely investigated in computer vision and natural language processing as a means to reduce the computational cost, memory footprint, and energy consumption of large language models \cite{xu2023survey, xu2025resource, gordon2020compressing, xu2021beyond, ye2019adversarial, du2021robustness, jiao2019tinybert, buciluǎ2006model, jiang2023lion, zhang2018structadmm, tang2019distilling}. Commonly adopted model compression techniques include pruning \cite{sanh2020movement}, quantization \cite{zafrir2019q8bert}, and knowledge distillation (KD) \cite{hinton2015distilling}. Recently, these techniques have been adapted for software analytics to improve the efficiency of code language models while maintaining acceptable performance \cite{shi2024efficient}. Wei et al. \cite{wei2023towards} examined the effects of quantization on code generation models and showed that, when applied appropriately, it can substantially reduce resource usage with only minor impacts on accuracy and robustness. Afrin et al. \cite{afrin2025quantization} investigated the effects of quantization beyond functional correctness, focusing on qualitative and static properties of generated code. Zhang et al. \cite{zhang2022diet} showed that streamlining input programs for CodeBERT improves efficiency with minimal impact on performance. Subsequently, Saad et al. \cite{saad2024alpine} proposed \textit{ALPINE}, a language agnostic pruning approach that reduces computational cost, while Aloisio et al. \cite{d2024compression} empirically compared pruning, quantization, and KD for CodeBERT across software analytics tasks. While pruning and quantization improve efficiency, their compression capacity is limited, as aggressive compression often degrades performance \cite{shi2022compressing}. Consequently, this study focuses on KD to obtain compact models while preserving predictive performance.

Shi et al. \cite{shi2022compressing} proposed \textit{Compressor}, a KD approach evaluated on CodeBERT and GraphCodeBERT for vulnerability prediction and clone detection, showing that model size and inference cost can be substantially reduced while preserving predictive performance. Building on this work, Shi et al. \cite{shi2024greening} introduced \textit{AVATAR}, which aims to reduce energy consumption and inference latency without sacrificing effectiveness. More recent studies have explored robustness-oriented and adaptive distillation strategies for code models. Panichella et al. \cite{panichella2025metamorphic} presented \textit{MORPH}, which integrates metamorphic testing with many objective optimizations to improve robustness on metamorphic code while jointly optimizing accuracy, efficiency, and model size. Liu et al. \cite{liu2026pioneer} proposed \textit{PIONEER}, which enhances the robustness of compressed code models with minimal accuracy loss across multiple architectures. Chen et al. \cite{chen2025smaller} introduced \textit{SODA}, a self-paced distillation framework that progressively transfers programming knowledge and outperforms prior distillation methods. Complementing these efforts, Wang et al. \cite{wang2025empirical} provided an empirical study showing that KD consistently outperforms finetuning for code understanding tasks, with feature-based approaches achieving the strongest performance.

In contrast to existing approaches that primarily optimize efficiency, robustness, or energy consumption in isolation, recent findings by Awal et al. \cite{awal2025metamorphic} demonstrate that even robustness-oriented KD methods such as \textit{MORPH} remain vulnerable to adversarial attacks. The findings indicate a gap in transferring decision-level knowledge through single-teacher distillation, limiting students’ capacity to learn effective and robust decision boundaries. To address this gap, we integrate a Mixture of Experts strategy into knowledge distillation to enable richer knowledge transfer and improve both predictive performance and adversarial robustness.


\section{Conclusion}
\label{end}

In this study, we introduce \textit{MoEKD}, a knowledge distillation framework that enhances both predictive performance and adversarial robustness in compressed language models of code, improving their usability and reliability. \textit{MoEKD} formulates distillation as a multi-teacher-to-student learning paradigm and leverages a Mixture-of-Experts (MoE) framework to aggregate specialized knowledge effectively. Our approach emphasizes the quality of transferred knowledge as a key factor in boosting the performance of compressed student models. Experimental evaluation on CodeBERT and GraphCodeBERT for the vulnerability detection task shows that \textit{MoEKD} improves predictive performance by up to \textbf{13.0\%} and adversarial robustness by up to \textbf{35.8\%} compared to state-of-the-art baselines, including \textit{Compressor} and \textit{AVATAR}. An ablation study confirms that aggregating high-quality knowledge from multiple experts is crucial for maintaining both accuracy and robustness in compact models. Future work will explore the generalizability of \textit{MoEKD} to additional models and software engineering tasks, and investigate advanced MoE architectures to further enhance expert knowledge aggregation for more effective knowledge distillation.

\bibliographystyle{ACM-Reference-Format}
\bibliography{sample-base}

\end{document}